 \documentstyle[12pt]{article}
\begin{document}
 \newcommand{\be}[1]{\begin{equation}\label{#1}}
 \newcommand{\ee}{\end{equation}}
 \newcommand{\beqn}[1]{\begin{eqnarray}\label{#1}}
 \newcommand{\eeqn}{\end{eqnarray}}
\newcommand{\mat}[4]{\left(\begin{array}{cc}{#1}&{#2}\\{#3}&{#4}\end{array}
\right)}
 \newcommand{\matr}[9]{\left(\begin{array}{ccc}{#1}&{#2}&{#3}\\{#4}&{#5}&{#6}\\
{#7}&{#8}&{#9}\end{array}\right)}
 \newcommand{\eps}{\varepsilon}
 \newcommand{\Ga}{\Gamma}
\newcommand{\s}{\sigma}
\newcommand{\D}{\Delta}
 \newcommand{\la}{\lambda}
\newcommand{\ov}{\overline}
\newcommand{\mucirc}{\stackrel{\circ}{\mu}}
\newcommand{\mast}{\stackrel{\ast}{m}}
\newcommand{\meps}{\stackrel{\circ}{\eps}}
\newcommand{\mcirc}{\stackrel{\circ}{m}}
\newcommand{\mcir}{\stackrel{\circ}{M}}
\newcommand{\geqsim}{\stackrel{>}{\sim}}
\renewcommand{\thefootnote}{\fnsymbol{footnote}}

\begin{titlepage}
\begin{flushright}
UMD-PP-96-34\\
September,1995
\end{flushright}
\vspace{10mm}

 \begin{center}
 {\Large \bf Upper Bound on the $W_R$ Mass in Automatically
R-conserving SUSY Models}\\

\vspace{1.3cm}
{\large R. Kuchimanchi$^a$ and R.N. Mohapatra$^b$ }
\\ [5mm]
$^a$ {\em Department of Physics,
University of Virginia, Charlottesville, VA 22901, U.S.A.}\\
$^b$ {\em Department of Physics, University of Maryland, College Park,
MD 20742, U.S.A.}
\end{center}

\vspace{2mm}
\begin{abstract}


We show that in automatically R-conserving minimal SUSY left-right
symmetric models there is a theoretical upper limit on the mass of
the right-handed $W_R$ boson given by $M_{W_R}\leq g M_{SUSY}/f$, where
$M_{SUSY}$ is the scale of supersymmetry breaking, g is the weak gauge
coupling and $f$ is the Yukawa coupling responsible
for generating the right-handed neutrino masses.
If $M_{W_R}$ violates the above limit, the
ground state of the theory breaks electromagnetism. The only way
to avoid this bound while keeping the theory automatically R-conserving
is to expand the theory to include very specific kinds of
additional multiplets and demanding unnatural finetuning of their couplings.
\end{abstract}

\end{titlepage}

\renewcommand{\thefootnote}{\arabic{footnote})}
\setcounter{footnote}{0}

\vspace{4mm}

There are two interesting extensions of the standard model which
are currently the focus of intensive investigation: the minimal
supersymmetric standard model (MSSM)\cite{hab84} and the left-right symmetric
model based on the gauge group $SU(2)_L\times SU(2)_R\times U(1)_{B-L}
\times SU(3)_c$\cite{lrs}. The MSSM extension can resolve several important
unanswered questions of the standard model: (i) it explains the
stability of the electroweak symmetry breaking scale under radiative
corrections from new physics at higher scales such as those due to
grand unification or gravity; (ii) local supersymmetry breaking
via renormalization group equations can provide a perturbative origin
for the scale of electroweak symmetry breaking; (iii) it may provide
a particle physics candidate for the cold dark matter of the Universe.
It also connects the standard model with more fundamental theories
such as superstrings. On the other hand, the left-right symmetric (LRS) models
(i) provide a more satisfactory framework for understanding the origin of
parity violation; (ii) restore quark-lepton symmetry to weak interactions
and (iii) if the neutrinos have a mass, the LRS models provide the simplest
way to understand the smallness of the neutrino mass via the see-saw mechanism
\cite{gel79}.

The next important question is the scale at which these new symmetries
manifest themselves. As far as supersymmetry is concerned, the general
belief is that its scale $M_{SUSY}$ is below or around a TeV if it has to
have the usefulness expected of it. On the other hand, as far as the
scale of left-right symmetry ( denoted by $M_{W_R}$ ) is concerned,
in general it could be anywhere
between a few hundred GeV to the GUT scale. The main result of this letter
is that within the theoretical framework that combines both supersymmetry
and left-right symmetry, there exist a class of attractive models
where one can derive an upper bound on $M_{W_R}$ related to $M_{SUSY}$.

In order to introduce the special class of models we discuss, we note that
despite its many attractive features, the MSSM extension has a major
drawback. While the standard model provides a natural understanding of why
baryon (B) and lepton (L) number conservation are obeyed to such a high degree
of precision in nature, in MSSM both B and L violation can occur with
maximal strength via the so called R-parity violating terms\cite{hal84}.
There is also no CDM candidate in the presence of these terms and the
usual practice is to impose an R-parity symmetry
( defined as $(-1)^{3B+L+2S}$ ) on the MSSM to avoid both these problems.
There is however a less ad hoc way to solve the problem of catastrophic
B-violation as has been already noted earlier\cite{mar92}
by extending the gauge group of the supersymmetric model to make it
left-right symmetric ( to be called SUSYLR ) and considering the
class of models which implement the see-saw mechanism for
small neutrino masses. Such models lead to automatic R-parity conservation
prior to symmetry breaking and thus have no problem with B or L
violation at all. This model has been
studied in several recent papers \cite{kuc93,cve84,fra91} .
Specifically, in our earlier paper \cite{kuc93}, we pointed out that
in the minimal version of this model, the  requirement of
electric charge conservation and low energy parity violation implies
that R-parity must be spontaneously broken \cite{aul}. Therefore,
while this model still does not have a CDM candidate, it cures one
major drawback of the MSSM i.e. the baryon number remains an exact
symmetry of the model. The model however leads to small
L-violating terms, which are suppressed and could
be tested experimentally by searches for L-violation.

In this letter, we report another interesting consequence of
this class of minimal SUSYLR models, which is that
the mass of the right-handed $W_R$
boson, $M_{W_R}$ has an upper limit related to the SUSY breaking scale
(i.e. $M_{W_R}\leq g M_{SUSY}/f$) where $g$ is the weak gauge coupling
and $f$ is the Yukawa couplings of the right handed neutrinos. The
Yukawa coupling $f$ could a priori
be of order one, in which case one gets the most stringent bound
on $M_{W_R}$ to be less than a TeV (since $M_{SUSY}$ is expected
to be in the TeV range). These results follow from
the requirements of electric charge conservation and low energy parity
violation by the ground state of the theory. The
$W_R$ in this class of models ( with $f\simeq 1$ ) then becomes
 detectible in high energy colliding machines such as the LHC.
It is worth pointing out that while there exist convincing arguments
based on analysis of the low energy weak processes which give
lower bounds on $M_{W_R}$ in the range of few hundred GeV, to
the best of our knowledge, there exist no upper bound on $M_{W_R}$.
We therefore feel that our result is interesting even though it
holds in a particular ( though realistic and interesting ) class
of models.
We further show that the only way to avoid this bound
while maintaining the feature of automatic R-parity conservation is to
enrich the minimal model by adding a $B-L=0$ triplet, whose Yukawa couplings to
$B-L = \pm 2$ triplets are unnaturally finetuned at the electroweak level.

\vspace{4mm}

We start our discussion by giving the matter content of the minimal
SUSYLR model: $Q(2, 1, {1 \over 3});$ $Q^c(1, 2, -{1 \over 3});$
$L(2, 1, -1);$ $L^c(1, 2, +1)$ denote the quarks and leptons and
$\D(3, 1, +2);$ $\D^c(1, 3, -2);$ $\overline{\D}(3, 1, -2);$ $\overline{\D^c}
(1, 3, +2);$ $\Phi(2, 2, 0)$ denote the Higgs fields ( where the numbers
in the parenthesis represent the representations of the fields under the
gauge group $SU(2)_L\times SU(2)_R\times U(1)_{B-L}$). Since
the weak interaction scale $M_{W}$ is smaller than $M_{SUSY}$ and
$M_{R}$, we will initially ignore effects of the order $M_{W}$.


For the sake of completeness, we start by presenting
the argument of reference\cite{kuc93}, regarding the inevitability
of spontaneous R-parity breaking in the SUSYLR models under
consideration if they have to provide a realistic
description of nature.  Let us write down the relevant part of the
superpotential. We have included a gauge singlet field $ ( \sigma ) $
to make the discussion as general as possible.

\begin{equation}
\label{eq:supers}
W =  i f \left ( {L}^{cT} \tau_{2} \D^{c} {L}^{c} \right )
+ M ~Tr \left ( \D^{c} \bar{\D}^{c} \right )
 + h_{\s}  Tr\left ( \D^{c} \bar{\D}^{c} \right ) \s + F( \s )
\end{equation}

The Higgs potential consists of a sum of the $F$ and $D$ terms ($V_{F}$ and
$V_{D}$ respectively) and the soft symmetry breaking terms ( we have dropped
all terms involving $\D$ and $\overline{\D}$ and $\Phi$
 since their vev's are at most of order $M_W$
 and will therefore not play any role in our discussion ).

\begin{equation}
V = V_{F} + V_{D} + V_{S}
\end{equation}
where
\begin{eqnarray}
\label{eq:fterms}
V_{F} = & ~& \left| 2 i f \tilde{L}^{cT} \tau_{2} \D^{c}\right|^{2}
+ \left| h_{\s} Tr \left ( \D^{c} \bar{\D}^{c} \right ) + {{\partial F} \over
{\partial \s}} \right|^{2}
\nonumber \\
& ~& + ~Tr \left| i f \tilde{L}^{c}
\tilde{L}^{cT} \tau_{2} + M \bar{\D}^{c} + h_{\s} \bar{\D}^{c} \s
\right|^{2} \nonumber \\
& ~& + ~Tr \left| M \D^{c} + h_{\s} \D^{c} \s \right|^{2},
\end{eqnarray}

\begin{eqnarray}
\label{eq:dterms}
V_{D} = & ~&+ {{g^{2}} \over 8}
\sum_{m} \mid \tilde{L}^{c\dagger} \tau_{m} \tilde{L}^{c}
+ Tr \left ( 2 \D^{c\dagger} \tau_{m} \D^{c}
+ 2 \overline{\D}^{c\dagger} \tau_{m} \overline{\D}^{c} \right ) \mid^{2}
\nonumber \\
& ~&+ {{g'^{2}} \over 8} \mid
\tilde{L}^{c\dagger} \tilde{L}^{c} -
2 Tr \left (   \D^{c\dagger} \D^{c}
- \bar{\D}^{c\dagger}\bar{\D}^{c}\right ) \mid^{2}
\end{eqnarray}

\begin{eqnarray}
\label{eq:vsoft}
V_{S} = & ~&m_{l}^{2} \tilde{L}^{c \dagger} \tilde{L}^{c} + M_{1}^{2}
Tr \D^{c \dagger} \D^{c}  + M_{2}^{2} Tr \bar{\D}^{c \dagger} \bar{\D}^{c}
+~Ah_{\sigma} Tr(\D^c\overline{\D^c}) \sigma \nonumber \\
& ~&+ \left ( M'^{2} Tr \D^{c} \bar{\D}^{c} + i f v
\tilde L^{cT} \tau_{2} \D^{c}
\tilde{L}^{c} + H.C. \right ) + F_{1}(\sigma)
\end{eqnarray}

All the fields in the above equations represent the scalar field of the
corresponding superfield.
This model will yield the MSSM for scales $\mu << \left < \D^{c}
\right >, \left < \bar{\D^{c}} \right >$; if $\left < \D^{c}
\right >, \left < \bar{\D^{c}} \right > \neq 0. $  R-parity will be
conserved or broken at this stage depending on
whether $ \left < \tilde{\nu^{c}}
\right > $ is zero or not.
We will now show that, if
$\left < \tilde{\nu^{c}} \right > = 0 $, then either
$\left < \D^{c} \right > = 0$ or if it is non-zero,
then the absolute global minimum of $V$ violates
electric charge ($Q_{em}$) conservation.

Note first that since in general $M_{1} \neq M_{2}$, $\left <\D^{c} \right >
\neq \left < \bar{\D^{c}} \right > $. In what follows, we will denote
the vev's of $\D^c$ and $\overline{\D^c}$ generically by $v_R$,
the right-handed symmetry breaking scale.
  Consider now the two vacua:

\noindent (a) \underline{ $Q_{em}$ conserving vev:}
\begin{equation}
\label{eq:cons}
\langle \D^{c} \rangle = d \left( \begin{array}{cc}
                      0 & 0 \\
                      1 & 0
                           \end{array}
                       \right),
\ ~~ \langle \overline{\D}^{c} \rangle = \overline{d}
   \left( \begin{array}{cc}
                      0 & 1\\
                      0 & 0
                           \end{array}
                       \right)
\end{equation}

\noindent (b) \underline{$Q_{em}$ breaking vev:}
\begin{equation}
\label{eq:break}
\langle \D^{c} \rangle = {{d} \over {\sqrt{2}}} \left( \begin{array}{cc}
                      0 & 1 \\
                      1 & 0
                           \end{array}
                       \right),
\ ~~ \langle \bar{\D}^{c} \rangle = {{\bar{d}} \over {\sqrt{2}}}
   \left( \begin{array}{cc}
                      0 & 1\\
                      1 & 0
                           \end{array}
                       \right)
\end{equation}

It is easy to see using properties of the Pauli matrices that if $\left <
\tilde{\nu^{c}}\right > = 0$, then the value of the positive definite
$V_{D}$ for the  $Q_{em}$ breaking vev is lower than $V_{D}$ for the
$Q_{em}$ conserving vev since $Tr \D^{c \dagger} \tau_{m} \D^{c} \neq 0$
in case (a) whereas it vanishes for case (b).  The value of all other terms
of $V$ are the same for both cases.
When R-parity is broken by giving a nonzero vev
$\langle \tilde{\nu^c} \rangle \equiv l'$,
there are new contributions to
 the $V$, and one can adjust parameters like $f$ to make the
$Q_{em}$ breaking vev to have a higher energy than the $Q_{em}$ conserving
one. It is however important that $l'$ be at least of order
$M_{SUSY}$ or the right-handed scale $v_R$ (whichever is lower) to
achieve this goal ( note that $d\sim \bar{d} \sim v_R$ ).

\vspace{4mm}
\noindent{\underline {\it Upper bound on $v_{R}$}}
\vspace{4mm}

Let us now proceed to the discussion of this paper.
We will now assume that
\begin{equation}
\label{eq:ass}
f v_{R} >> M_{SUSY}
\end{equation}
 and hence to the lowest
order in this approximation we can neglect the soft supersymmetry breaking
terms, $V_{S}$.
Neglecting these terms the minima are SUSY preserving and they satisfy
the condition $V = 0$.  This requires that
\begin{equation}
\label{eq:susypres}
\tilde{L}^{c} = 0, ~~ M + h_{\s} \s = 0,
{}~~Tr \left ( \D^{c} \bar{\D}^{c} \right ) = - {{\partial F} \over
{h_{\s}{\partial \s}}}, ~~~
Tr \D^{c \dagger} \D^{c} = Tr \bar{\D}^{c \dagger} \bar{\D}^{c}
\end{equation}

It is interesting to note that at this stage $\tilde{L}^{c} = 0$ follows
from the SUSY preserving condition $V = 0$ and is not put in by hand.
In other words, to the extent that supersymmetry is not broken,
R-parity cannot be spontaneously broken.

The other more important thing to note is that
the solutions to  equation~\ref{eq:susypres} give rise to degenerate minima.
The solutions are:
\begin{equation}
\label{eq:diffdels}
\left < \D^{c} \right > = d \left ( \begin{array}{cc}
                      0 & sin \theta \\
                     cos \theta & 0
                           \end{array}
                       \right ),
\ ~~ \left < \bar{\D}^{c} \right > = \bar{d}   \left ( \begin{array}{cc}
                      0 & cos \theta \\
                     sin \theta & 0
                           \end{array}
                       \right ),
\end{equation}
and
\begin{equation}
\label{eq:sigma}
\ ~~ \left < \s \right > = {{-M} \over h_{\s}}, ~~ \left <
\tilde{L}^{c} \right > = 0
\end{equation}
with
\begin{equation}
\label{eq:d=d'}
\bar{d}^{2} = {d }^{2} = - {{\partial F} \over {\partial \s}}.
\end{equation}
In equation~(\ref{eq:diffdels}),
$\theta$ can be  any angle, and this corresponds to the
degeneracy of the vacua.

In order to lift the degeneracy, and see which of the above vacua is the
true minimum, we turn on the
most general soft supersymmetry breaking terms ( $V_{Soft}$ )
as a perturbation.  $V_{soft}$ is given by equation~(\ref{eq:vsoft}) with
 $m_{l}, v, M_{1}, M_{2}, M'$ are all at most of the order
$M_{SUSY} << v_{R}$.

Note that as already mentioned, since $M_{W} < M_{SUSY}$, we set
$\left < \Phi \right > = \left < \tilde{L} \right > = M_{Weak} = 0$
in the lowest order approximation.  Apriori, $\left < \tilde{L}^c
\right > $ can be order $M_{SUSY}$.  However we will show in the
following that this cannot be the case, and that in fact
$\tilde{L}^{c} = 0$ to the lowest order.

To see if $\tilde{L}^{c}$ picks up a VEV, let us use
equation~(\ref{eq:diffdels}) with $ \theta = 0$ and $d \sim \bar{d} \sim
v_R$
and examine the $\tilde{L}^c$ terms.
Substituting for $\D^{c}, \bar{\D}^{c}$ and $l'$ into
equations~(\ref{eq:fterms}),~(\ref{eq:dterms})~and~(\ref{eq:vsoft}) we have
the
following quadratic terms in $l'$
\begin{eqnarray}
\label{eq:lterms}
\left| 2 i f \tilde{L}^{c T} \tau_{2} \D^{c} \right|^{2} \sim  ~ \left| f
\right|^{2} v_{R}^{2} l'^{2}; \nonumber \\
Tr \left| i f \tilde{L}^{c} \tilde{L}^{c T} \tau_{2} + \left ( M +
h_{\s} \s \right ) \bar{\D}^{c} \right|^{2} \sim  ~f v_{R} M_{SUSY}
l'^{2}; \nonumber \\
f v \tilde{L}^{c T} \tau_{2} \D^{c} \tilde{L}^{c} \sim  ~ f M_{SUSY} v_{R}
l'^{2}; \nonumber \\
m_{l^c}^{2} \tilde{L}^{c \dagger} \tilde{L}^{c} \sim ~ M_{SUSY}^{2}
l'^{2}; \nonumber \\
{{g'^{2}} \over 8} \left| \tilde{L}^{c \dagger} \tilde{L}^{c} +
2 Tr \left ( \bar{\D}^{c}  \bar{\D}^{c} - \D^{c \dagger} \D^{c} \right)
\right|^{2} \sim  ~ \left ( g'^{2} ~ or ~ g^{2}\right ) M_{SUSY}^{2} l'^{2}
\end{eqnarray}
It is important to note that there are no linear or cubic terms in
$\tilde{L}^c$ and the rest are all quartic terms.  Thus assuming
inequality~(\ref{eq:ass}) we have the following relation from
the above equations.
\begin{equation}
\label{eq:l'=0}
V_{\tilde{L}^{c}} \approx |f v_{R}|^{2} |l'|^{2} + ~\mu^2~ |l'|^{2} +
{}~const.~|l'|^{4}
\end{equation}
where $\mu$ denotes a mass of the order of $M_{SUSY}$.
It is easy to see that at the minimum $l' = 0$ or $\tilde{L}^{c} =
0$.  This means that there cannot be any spontaneous R-parity
violation if inequality~(\ref{eq:ass}) is true.  But from our result
given earlier, this means that electromagnetism is spontaneously broken.
Hence inequality~(\ref{eq:ass}) must be false and we have the bound
\begin{equation}
\label{eq:bound1}
v_{R} \le {{M_{SUSY}} \over f}
\end{equation}
which implies a bound on $M_{W_{R}} \le g M_{SUSY} / f $.

Note that if $fv_R\leq M_{SUSY}$, then the first two terms in Eq.(14)
would be of same order and could lead to a net negative value
for the $(mass)^2$ term for the field $l'$ and thereby give a non-zero vev
for $l'$ ( causing R-parity violation and therefore a $Q_{em}$ conserving
and parity violating low energy theory ).

A way of understanding the role played by $V_{S}$ in preferring the
electro-magnetism violating vacuum is to note that because in general
$M_{1} \neq M_{2}$ in equation~(\ref{eq:vsoft}), the magnitude of the
VEVs of $\D^{c}$ and $\bar{\D}^{c}$ are slightly different.  Now the
degenerate vacuua of equation~(\ref{eq:diffdels}) will be slightly split.
Substituting equations~(\ref{eq:diffdels}) and (\ref{eq:d=d'}) into
equations~(\ref{eq:fterms}),~(\ref{eq:dterms}) and (\ref{eq:vsoft})
 it is easy to see that the minimum value is obtained for $\theta =
45^{0}$
since $d \neq \bar{d}$.  This vacuum violates electromagnetism.

The upperbound in equation~(\ref{eq:bound1}) implies that $v_{R} \le
M_{SUSY}$ if the yukawa couplings in the right-handed sector are of
the order 1. This is the main result of our paper.
Since the bound on $v_R$ depends on the unknown Yukawa
coupling $f$, it gets weaker as $f$ becomes much smaller than one.

So far we have ignored $M_{W}$ and hence
have not considered the effect of the VEV of the bidoublet.
  We will now show that turning on
$\left < \Phi \right > \sim M_{W}$ and $\left < \tilde{L} \right >
\le M_{W}$ doesn't change
any of the conclusions arrived in the previos subsections.
Due to terms like $\left| h \tilde{L}^{T} \tau_{2} \phi \tau_{2} +
2 i f \tilde{L}^{c T} \tau_{2} \D^{c} \right|^{2}$
equation~(\ref{eq:l'=0}) will be modified  to
\begin{equation}
\label{eq:l'=0+}
V_{\tilde{L}^{c}} \approx f M_{W}^{2}v_{R}l' + |f v_{R}|^{2} |l'|^{2}
+ ~\mu^2~ |l'|^{2} +~const.~|l'|^{4}
\end{equation}
Note that in this case the effective $(mass)^2$ term for the $l'$
is still positive.
At the minimum, $l' \sim M_{W} {{M_{W}} \over {f v_{R}}} < < M_{W}$,
if $f v_{R} > M_{SUSY}$.
This value of $\langle l' \rangle$ is much too small to stabilize the
electromagnetism conserving vacuum, and therefore the bound
$f  v_{R} < M_{SUSY}$ holds even after electroweak effects
are taken into account.  In fact,
the splitting in the magnitudes of $\D^{c}$
and $\bar{\D}^{c}$ caused by this tiny nonzero $l'$
causes the theory to prefer the electro-magnetism violating vacuum
if the bound is violated.

\vspace{4mm}
\noindent{\underline{\it SUSYLR with $B-L=0$ triplet as a way to
avoid the bound:}}
\vspace{4mm}

Let us now discuss what one has to do avoid the above bound on $m_{W_R}$
while at the same time maintaining automatic R-parity conservation in
the theory prior to symmetry breaking. For this purpose,
let us enrich the minimal models by adding
a $B-L=0$ triplet denoted by $\omega$.
We add to the fields given earlier a $B-L = 0$, $SU(2)_{R}$
triplet field $( \omega ) $ and a singlet field $\s$
already included in the model.
The relevant part of the Superpotential is given by
\begin{eqnarray}
\label{eq:b-lsuper}
W = & ~& M Tr \left ( \D^{c} \bar{\D}^{c} \right ) + a Tr \left (
\D^{c} \omega \bar{\D}^{c} \right ) +  b Tr \left (
\bar{\D}^{c} \omega {\D}^{c} \right ) \nonumber \\
 & ~& +~c Tr \omega^{2} + h_{\s} Tr \left ( \D^{c} \bar{\D}^{c} \right ) \s
+ h_{\omega} Tr \omega^{2} \s + f \left ( \s \right )
\end{eqnarray}
Note that in the above $\omega$ and $\D^{c}$ do not commute with each other.
Differentiating the above with respect to the complex field variables
we get the following F-term
\begin{eqnarray}
\label{eq:b-lfterm}
V_{F} = & ~& Tr \left|M \D^{c} + a \D^{c} \omega + b \omega \D^{c} + h_{\s} \s
\D^{c} \right|^{2} \nonumber \\
 & ~& + Tr \left| M \bar{\D}^{c} + a \omega \bar{\D}^{c} + b \bar{\D}^{c}
\omega
+ h_{\s} \s \bar{\D}^{c} \right|^{2} \nonumber \\
 & ~& + Tr \left| a \bar{\D}^{c} \D^{c} + b \D^{c} \bar{\D}^{c} + 2 c
\omega + 2 h_{\omega} \s \omega \right|^{2} \nonumber \\
 & ~& + Tr \left| h_{\s} Tr \left ( \D^{c} \bar{\D}^{c} \right )+
 h_{\omega} Tr
\omega^{2} + {{\partial f} \over {\partial \s}} \right |^{2}
\end{eqnarray}

We will look for supersymmetry preserving minima, since we are
interested in seeing if we can violate the upper bound on $v_{R}$ we
derived.  Using $SU(2)_R$ invariance, we
can always choose a basis such that $\omega$ acquires the
electro-magnetism preserving
vaccum expectation value of the form
\begin{equation}
\label{eq:o}
\left < \omega \right >  = w \left ( \begin{array}{cc}
                      1 & 0 \\
                      0 & - 1
                           \end{array}
                       \right ).
\end{equation}
By substituting vacuum expectation values for $\D^{c}$ and $\bar{\D}^{c}$
from equation~(\ref{eq:diffdels}) into equation~(\ref{eq:b-lsuper}) and using
equation~(\ref{eq:o}), we see that the superpotential W has a non-trivial
$\theta$ dependence due to the trilinear terms involving the three
triplet fields.  Since the soft-supersymmetry breaking terms
will have exactly the same form as the superpotential, the value of the Higgs
potential will have non-trivial $\theta$ dependence even if R-Parity
is not spontaneously broken.  Thus there is no problem of degenerate vacuaa
and these trilinear couplings can have signs so that the electromagnetism
conserving vacuum with $\theta = 0$ is an absolute minimum even without
R-Parity violation.  In the following we will therefore assume that
the rest of the triplet fields have $Q_{em}$ conserving
vaccum expectation values
as in equation~(\ref{eq:diffdels}) with $\theta = 0$.

The supersymmetry conserving minima should have $V_{F} = 0$ and
substituting
this in equation~(\ref{eq:b-lfterm}) gives the following equations:
\begin{equation}
\label{eq:a}
a  d \bar{d} + 2 c w + 2 h_{\omega} w \s = 0 \nonumber
\end{equation}
\begin{equation}
\label{eq:b}
b d \bar{d} - 2 c w - 2 h_{\omega} w \s = 0 \nonumber
\end{equation}
\begin{equation}
M + (a - b) w + h_{\s} \s = 0 \nonumber
\end{equation}
\begin{equation}
\label{eq:problem}
h_{\s} d \bar{d} + 2 h_{\omega} w^{2} + {{\partial f} \over {\partial
\s}} = 0
\end{equation}

Note that the two equations~(\ref{eq:a}) and (\ref{eq:b}) follow
from the third term in equation~(\ref{eq:b-lfterm}).
We see that these two relations cannot be simultaneously satisfied
unless
\begin{equation}
\label{eq:a=-b}
a = - b.
\end{equation}

Thus the two coupling parameters $a$ and $b$ must be fine-tuned to be
equal to the order
$M_{SUSY} / v_{R}$ since $V_{F}$ can at best pick up a small value due
to SUSY violating soft-symmetry breaking terms.
Such a fine-tuning is highly unpleasing unless it is forced by some
symmetry.  There is no obvious symmetry at the level of the left-right
model that can lead to equation~(\ref{eq:a=-b}).
\vspace{4mm}

In conclusion, we have studied the low energy implications of the class
of SUSY left-right models that lead to automatic R-parity conservation.
The most important result of our investigation is that
 in the minimal version of this
model, there is an upper bound on $M_{W_{R}}$, which for allowed values
of parameters in the theory can be as small as a TeV making it
possible to rule out the model in this parameter range.

\vspace{4mm}
 We like to thank A. Rasin for carefully reading the manuscript
and for useful comments.
The work of R. N. M. was supported by a grant
from the National science Foundation under a grant no. PHY-9119745
and by a Distinguished Faculty Research Award by the University of
Maryland for the year 1995-96. The work of R.K. was supported
in part by the U.S. Department of Energy under grant No. DE-AS05-89ER40518.

\end{document}